\documentclass{l4dc2027}
\usepackage[T1]{fontenc}
\usepackage{times}
\usepackage{microtype}
\usepackage{mathtools}
\usepackage{booktabs}
\usepackage{enumitem}

\usepackage{xcolor}

\newcounter{vipcounter}

\newcommand{\R}{\mathbb R}
\newcommand{\E}{\mathbb E}
\newcommand{\tr}{\operatorname{tr}}

\newtheorem{assumption}[theorem]{Assumption}

\hypersetup{colorlinks=true,linkcolor=blue,citecolor=blue,urlcolor=blue}

\title[First-Order Model-Free Policy Gradient for Undiscounted LQR]{First-Order Model-Free Policy Gradient for Undiscounted LQR \\with Global Optimality and Convergence Guarantees}
\coltauthor{%
 \Name{Vipul K Sharma} \Email{sharm697@purdue.edu}\\
 \Name{S Sivaranjani} \Email{sseetha@purdue.edu}\\
 \addr Edwardson School of Industrial Engineering, Purdue University,
 West Lafayette, IN 47906, USA%
}
\begin{document}
\raggedbottom
\maketitle
\begin{abstract}
We provide a first-order model-free policy gradient (PG) framework for learning undiscounted infinite-horizon linear quadratic regulator (LQR) policies with global optimality and convergence guarantees. We first construct a stochastic wrapper policy class parameterized by the LQR gain to enable exploration around a deterministic linear control law. We analyze the additional cost induced by exploration over infinite-horizon, and establish a weak gradient dominance condition that characterizes global optimality under exploration. We then construct a model-free first-order PG algorithm using sampled trajectory information, without identifying the dynamics or approximating a value function. Finally, we establish finite-time convergence guarantees, providing the sufficient number of samples required for near-global optimality with our proposed PG algorithms.
\end{abstract}
\section{Introduction}
The linear quadratic regulator (LQR) is a canonical optimal control problem for linear dynamical systems, whose optimal policy belongs to the class of stabilizing linear state-feedback controllers \citep{anderson2007}. When the dynamics are known, the optimal gain can be computed by solving the algebraic Riccati equation. When the dynamics are unknown, a standard approach is to identify the system matrices from data and then synthesize a controller using the estimated model \citep{dean2020}. Both approaches require an explicit model for controller synthesis. In contrast, reinforcement learning (RL) provides a model-free framework where the controller is learned directly from interactions with environment, without a separate system-identification step \citep{recht2018}.

When the policy structure is known, as in LQR, controller learning can be formulated as policy optimization over the parameters of the structured policy class. In the model-based setting, \citet{fazel2018} showed that exact policy gradient (PG) updates converge globally to an optimal LQR controller despite the nonconvex landscape, by establishing gradient dominance. Subsequent work has related gradient dominance to hidden convexity, providing a unified analysis for LQR \citep{watanabe2026revisiting,watanabe2026unified}. More broadly,
\citet{bhandari2024global} connect policy improvement to global optimality
of stationary policies, with applications to discounted LQR. In the model-free setting, \citep{fazel2018} adopts a zeroth-order approach, estimating the gradients by perturbing the controller gain and evaluating its corresponding objective. Related zeroth-order methods provide convergence and
sample-complexity guarantees for LQR \citep{malik2019,mohammadi2020} and
for distributed linear-quadratic control \citep{furieri2020}.

In such zeroth-order methods, sampling noise can mask the objective
change caused by a small gain perturbation, yielding noisy gradient estimates. This motivates first-order approaches that exploit the differentiable policy structure to estimate gradients from trajectory data. Model-free actor-critic methods provide one such approach using a learned value function \citep{zhou2023,ju2025}. However, they require an additional policy-evaluation procedure, and critic estimation errors affect the policy updates even when the quadratic value-function structure is known. We therefore study a direct model-free first-order PG construction for LQR using sampled trajectories and accumulated costs, without system identification or value-function approximation.

We develop this first-order PG construction by introducing a stochastic wrapper around a linear feedback controller. This stochastic wrapper facilitates exploration for model-free learning with a deterministic LQR policy. 
This construction builds on stochastic-policy approaches to learning deterministic controllers \citep{montenegro2024}, including structured policies such as PID and constrained LQR \citep{sharma2025pid,sharma2026safe}. However, the undiscounted infinite-horizon setting introduces a critical challenge. Exploration affects both the current input and, through the dynamics, subsequent states and feedback inputs; as a result, persistent exploration can make the expected cumulative cost infinite even when the underlying linear controller is stabilizing. 
To address this, we introduce decaying exploration and characterize the resulting infinite-horizon objective. Since the exploration cost depends on the gain, finiteness alone does not establish whether optimizing this stochastic objective recovers the optimal classical LQR controller. We show that the stochastic and deterministic objectives share the same unique optimal gain and establish weak gradient dominance for the stochastic objective. 
Using this relation, we establish finite-time convergence and sample-complexity guarantees for model-free PG, with the effect of exploration explicitly characterized in the performance bound. Our main contributions are as follows:
\begin{itemize}
 \item \textbf{Learning model-free LQR with exploration.} 
 We construct a stochastic wrapper policy centered around the deterministic linear feedback controller allowing exploration by adding a Gaussian noise constructed so as to ensure that infinite-horizon undiscounted objective remains bounded under such stochastic exploration. This allows us to apply the classic PG theorem \citep{sutton1999} using stochastic policies during training while learning the optimal underlying deterministic LQR controller.
 \item \textbf{First-order model-free policy gradient with finite-time convergence and global optimality.} We derive the exact score functions for the structure stochastic wrapper policy and construct a first-order PG update using sampled states, control inputs, and accumulated costs. We establish weak gradient dominance for the stochastic objective and derive finite-time performance and sample-complexity guarantees for model-free PG, explicitly characterizing the effect of exploration on the performance bounds.
\end{itemize}
\textbf{Organization.}
We first formulate the model-free LQR problem in Section~\ref{sec:problem}. Then, in
Section~\ref{sec:wrapper}, we introduce the stochastic wrapper policy to enable exploration and
show that this exploration preserves the optimal LQR gain.
Further, we develop the model-free policy-gradient
algorithm in Section~\ref{sec:algorithm}, and establish global optimality and finite-time convergence results for the proposed algorithm in Section~\ref{sec:convergence}. Finally, we provide the proofs for our results in the Appendix~\ref{app:proofs} with additional details in the supplementary material.\\
\textbf{Notation.} 
We denote the set of real numbers by $\R$. The norm $\|\cdot\|$ denotes the Euclidean norm for vectors and the induced spectral norm for matrices, while $\|\cdot\|_F$ denotes the Frobenius norm. Gradients with respect to matrix parameters are defined using the Frobenius inner product $\langle U,V\rangle_F=\tr(U^\top V)$. For a symmetric matrix $M$, the relations $M\succeq0$ and $M\succ0$ denote positive semidefiniteness and positive definiteness, respectively, and $\lambda_{\min}(M)$ denotes its smallest eigenvalue. We write $\rho(M)$ for the spectral radius of a square matrix $M$. For $a\in\R$, we define $(a)_+=\max\{a,0\}$.

\section{Problem Formulation}\label{sec:problem}
We consider the discrete-time linear system
\begin{equation}\label{eq:dynamics}
 x_{t+1}=Ax_t+Bu_t,\qquad x_0\sim\mathcal D,
\end{equation}
where $x_t\in\R^n$ and $u_t\in\R^m$ denote the state and control input, respectively, $\mathcal D$ represents initial-state distribution, and $A\in\R^{n\times n}$ and $B\in\R^{n\times m}$ are the system and input matrices.
Throughout, we assume that matrices $A$ and $B$ are unknown, and that we only have access to observed states, applied inputs, and stage costs $c_t:=c(x_t,u_t)$ at time $t$, where
$c(x,u):=x^\top Qx+u^\top Ru$ and $Q\succ0$ and $R\succ0$ are the state
and input cost matrices, respectively.
%

We now present the classical LQR problem, that seeks a causal control sequence or law solving the following undiscounted infinite-horizon problem
\begin{equation}\label{eq:classical-lqr}
 \begin{aligned}
 \underset{\{u_t\}_{t\ge0}}{\operatorname{minimize}}\quad&
 \E\!\left[\sum_{t=0}^{\infty}c(x_t,u_t)\right]\\
 \operatorname{subject\ to}\quad&
 x_{t+1}=Ax_t+Bu_t,\qquad x_0\sim\mathcal D,
 \end{aligned}
\end{equation}
where each $u_t$ may depend on the observed state history. We assume that $(A,B)$ is stabilizable. Classical LQR theory then establishes that the optimal control law is a stabilizing linear state-feedback policy \citep{anderson2007}, which motivates the gain parameterization $\mu_K(x)=Kx$, with $K\in\R^{m\times n}$. The corresponding closed-loop matrix is $F_K:=A+BK$, and the set of stabilizing gains is $\mathcal S:=\{K\in\R^{m\times n}:\rho(F_K)<1\}$. We impose the following condition on the initial-state.
\begin{assumption}\label{as:plant}
The initial state satisfies $\E[\|x_0\|^2]<\infty$ and $\Sigma_0:=\E[x_0x_0^\top]\succ0$.
\end{assumption}
This assumption is standard in LQR policy optimization \citep{fazel2018}.
In particular, $\Sigma_0\succ0$ ensures that the possible initial states span the entire state-space, so controller performance is sufficiently evaluated.
%
%
Equivalently, we can reformulate problem~\eqref{eq:classical-lqr} in terms of gain $K$ as
\begin{equation}\label{eq:det-problem}
 \min_{K\in\mathcal S}J_\mu(K),\qquad
 J_\mu(K):=\E\!\left[\sum_{t=0}^{\infty}c(x_t,Kx_t)\right],
\end{equation}
where $x_{t+1}=F_Kx_t$ and $x_0\sim\mathcal D$. 
Under Assumption~\ref{as:plant} and $Q\succ0$, it follows that $J_\mu(K)<\infty$ iff $K\in\mathcal S$. Thus, finite undiscounted objective enforces closed-loop stability within the linear feedback policy class. 
We denote the unique optimal gain by $K^*$ and the optimal value by $J_\mu^*:=J_\mu(K^*)$.

Our goal is to learn the optimal LQR gain $K^*$ without knowledge of the
system dynamics. To this end, we use a stochastic wrapper policy centered
at the deterministic linear feedback controller $\mu_K(x)=Kx$ during training
and construct model-free policy-gradient updates for $K$ from the
resulting trajectories, as presented in the following section.


\section{Exploration with LQR Policies}\label{sec:wrapper}
We first construct a stochastic wrapper policy class that enables exploration around linear feedback controllers. We then characterize the resulting infinite-horizon LQR objective with exploration and establish that optimizing over this class preserves the optimal classical LQR gain.
\subsection{Stochastic Wrapper Policy}
To explore around the deterministic linear controller $\mu_K(x_t)=Kx_t$, we define the stochastic wrapper policy, at time $t$, as
\begin{equation}\label{eq:wrapper-density}
 \pi_K(\cdot\mid x_t)=\mathcal N(\mu_K(x_t),\sigma_t^2\Omega),
 \qquad \sigma_t:=\frac{\sigma}{t+1},\quad t\ge0,
\end{equation}
where $\Omega\in\R^{m\times m}$, $\Omega\succ0$, and $\sigma>0$ are fixed independently of $K$. 
Equivalently, any sampled input $u_t \sim \pi_K(\cdot\mid x_t)$ can be written as
\begin{equation}\label{eq:wrapper}
 u_t=Kx_t+\sigma_tz_t,\qquad z_t\sim\mathcal N(0,\Omega),
\end{equation}
where the variables $z_t$ are independent across time and independent of $x_0$. Consequently, $\E[u_t\mid x_t]=Kx_t$, so the gain $K$ parameterizes the conditional mean of the stochastic policy.

The exploration noise $\varepsilon_t:=\sigma_tz_t$ affects both the current input and subsequent states through the closed-loop dynamics. Since $\E[u_t^\top Ru_t]\ge\sigma_t^2\tr(R\Omega)$, a constant positive exploration covariance would yield an infinite undiscounted LQR objective, even for a stabilizing gain. The decaying covariance in \eqref{eq:wrapper-density} controls this accumulation by ensuring
$\sum_{t=0}^{\infty}\E[\|\varepsilon_t\|^2]
=s_\sigma\tr(\Omega)<\infty$,
where 
\begin{equation}\label{eq:s_sigma}
    s_\sigma:=\sum_{t=0}^{\infty}\sigma_t^2=\pi^2\sigma^2/6.
\end{equation}
Together with stability of $F_K$, this summability ensures a finite expected exploration cost. For fixed $\sigma$ and $\Omega$, we define the stochastic wrapper policy class $\mathcal P_{\sigma,\Omega}:=\{\pi_K:K\in\mathcal S\}$,
with $\pi_K$ from~\eqref{eq:wrapper-density}. 

\subsection{LQR Objective with Exploration}
We formulate optimization over $\mathcal P_{\sigma,\Omega}$ as the following problem in the feedback gain $K$:
\begin{equation}\label{eq:stochastic-problem}
 \min_{K\in\mathcal S}J_\pi(K),\qquad
 J_\pi(K):=\E_{\pi_K}\!\left[\sum_{t=0}^{\infty}c(x_t,u_t)\right],
\end{equation}
where trajectories start from $x_0\sim\mathcal D$, evolve according to \eqref{eq:dynamics}, and use inputs generated by \eqref{eq:wrapper}. The expectation is taken over the initial state and exploration noise. To compare $J_\pi$ with $J_\mu$, we use the classical LQR policy-evaluation result: for any stabilizing gain $K\in\mathcal S$, the cost-to-go under $\mu_K$ is quadratic, $V_K(x)=x^\top P_Kx$ \citep{anderson2007}, where $P_K\succ0$ is the unique solution of
\begin{equation}\label{eq:lyap}
 P_K=Q+K^\top RK+F_K^\top P_KF_K.
\end{equation}
Taking expectation over $x_0\sim\mathcal D$ therefore gives $J_\mu(K)=\E[V_K(x_0)]=\tr(P_K\Sigma_0)$.
The following result establishes regularity of $J_\pi$ and explicitly characterizes its accumulated exploration cost.

\begin{theorem}[Cost of Exploration]\label{thm:cost}
Suppose Assumption~\ref{as:plant} holds. Then, $J_\pi$ is finite and twice continuously differentiable on $\mathcal S$. Furthermore, for every $K\in\mathcal S$,
\begin{equation}\label{eq:cost}
 J_\pi(K)=J_\mu(K)+\Delta(K),\qquad
 \Delta(K):=s_\sigma\tr\!\big((R+B^\top P_KB)\Omega\big).
\end{equation}
\end{theorem}
In \eqref{eq:cost}, the term $s_\sigma\tr(R\Omega)$ is the expected accumulated cost corresponding to $\varepsilon_t^\top R\varepsilon_t$, with $s_\sigma$ from \eqref{eq:s_sigma}. The same noise perturbs the next state by
$B\varepsilon_t$. As these perturbations propagate through the closed
loop, they affect both the states and the feedback inputs, contributing
the remaining accumulated cost $s_\sigma\tr(B^\top P_KB\Omega)$. Suppose
$W:=\Sigma_0+s_\sigma B\Omega B^\top\succ0$, we can then write
\begin{equation}\label{eq:weighted-cost}
 J_\pi(K)=\tr(P_KW)+s_\sigma\tr(R\Omega).
\end{equation}
Relative to $J_\mu(K)=\tr(P_K\Sigma_0)$, exploration replaces
$\Sigma_0$ by $W$ and adds the
offset $s_\sigma\tr(R\Omega)$.

\subsection{Optimal Gain Preservation}
We note that finiteness of the above exploration cost does not alone establish that optimizing $J_\pi$ recovers $K^*$, since $\Delta(K)$ depends on the gain. Using \eqref{eq:weighted-cost}, the following result establishes this property and compares the corresponding optimality gaps. 
\begin{theorem}\label{thm:optimality}
Suppose Assumption~\ref{as:plant} holds and $J_\pi^*:=\inf_{K\in\mathcal S}J_\pi(K)$. Then, $K^*$ is the unique minimizer of $J_\pi$ over $\mathcal S$, and $J_\pi^*=J_\mu^*+\Delta(K^*)$. Furthermore, for every $K\in\mathcal S$, we have
\begin{equation}\label{eq:deploy}
 J_\mu(K)-J_\mu^*
 \le J_\pi(K)-J_\pi^*.
\end{equation}
\end{theorem}
Theorem~\ref{thm:optimality} shows that exploration preserves the optimal gain. Moreover, \eqref{eq:deploy} ensures that a gain achieving nearly optimal cost under the stochastic policy also yields a nearly optimal deterministic linear controller. We can therefore learn the desired deterministic controller by optimizing $J_\pi$. We next construct the corresponding model-free PG update using sampled states, inputs, and costs.
\section{Model-Free LQR Policy Gradient}\label{sec:algorithm}
We now construct a model-free policy-gradient for $J_\pi$ and update the feedback gain $K$, using the stochastic wrapper policy score and sampled accumulated costs.
%
%
\subsection{Score and Gradient Estimator}
For fixed $x_t$ and $u_t$, the log density in
\eqref{eq:wrapper-density} depends on $K$ through the quadratic term
$-(u_t-Kx_t)^\top\Omega^{-1}(u_t-Kx_t)/(2\sigma_t^2)$.
Differentiating this term gives the policy score
\begin{equation}\label{eq:score}
 q_t(K):=\nabla_K\log\pi_K(u_t\mid x_t)
 =\frac{\Omega^{-1}(u_t-Kx_t)x_t^\top}{\sigma_t^2}.
\end{equation}

Further, let $\tau_H=(x_0,u_0,c_0,\ldots,x_{H-1},u_{H-1},c_{H-1},x_H)$ denote a trajectory of length $H\ge1$ generated under $\pi_K$ from \ref{eq:wrapper-density}.
Then, following \citet{williams1992,sutton1999}, we weight each policy score
$q_t(K)$ by the sampled cost-to-go $C_{t,H}:=\sum_{j=t}^{H-1}c_j$
and sum the resulting terms from $t=0$ to $H-1$ to obtain the Monte Carlo
policy gradient estimate
\begin{equation}\label{eq:finite-score-sample}
 G_H(K;\tau_H):=\sum_{t=0}^{H-1}q_t(K)C_{t,H}.
\end{equation}
Both the score and the cost-to-go are computed only from observed states and 
inputs following the stochastic wrapper policy, making the gradient estimator model-free.
%
\subsection{Policy Gradient Algorithm}
Let $N\ge1$ denote the batch size and $H\ge1$ the rollout length. We use $k$ to index PG iterations, $i=1,\ldots,N$ to index rollouts within an iteration, and $t=0,\ldots,H-1$ to index time steps within a rollout. At iteration $k$, we first collect $N$ trajectories $\{\tau_H^{(i)}\}_{i=1}^N$ under $\pi_{K_k}$. 
For each trajectory, we draw the initial state independently from
$\mathcal D$ and sample fresh exploration noise at each time step.
The time index $t$ in \eqref{eq:wrapper-density} is local to each
trajectory, so the exploration scale is $\sigma_t=\sigma/(t+1)$ for
$t=0,\ldots,H-1$ and starts at $\sigma_0=\sigma$ for every trajectory.
Finally, we average the resulting trajectory gradient estimates and use their
mean to update the LQR gain as
%
\begin{equation}\label{eq:batch-update}
 \widehat g_k:=\frac1N\sum_{i=1}^{N}G_H(K_k;\tau_H^{(i)}),
 \qquad K_{k+1}=K_k-\zeta\widehat g_k,
\end{equation}
where $\zeta>0$ is the step size. For rollout $i$, the sampled costs-to-go are computed backward using $C_{H,H}^{(i)}=0$ and $C_{t,H}^{(i)}=c_t^{(i)}+C_{t+1,H}^{(i)}$. Algorithm~\ref{alg:pg} summarizes the procedure over $T$ PG iterations.

\begin{algorithm2e}[h]
\caption{Model-Free Policy Gradient for Undiscounted LQR}\label{alg:pg}
\KwIn{$K_0\in\mathcal S$, $\sigma$ ,$\Omega$, $H$, $N$, $\zeta$, $T$.}
\For{$k=0,\ldots,T-1$}{
 \For{$i=1,\ldots,N$}{
  Sample $x_0^{(i)}\sim\mathcal D$\;
  \For{$t=0,\ldots,H-1$}{
   Sample $u_t^{(i)}\sim\pi_{K_k}(\cdot\mid x_t^{(i)})$ according to
\eqref{eq:wrapper-density}\;
   Apply $u_t^{(i)}$ and observe $c_t^{(i)}$ and $x_{t+1}^{(i)}$\;
  }
  Compute $G_H(K_k;\tau_H^{(i)})$ using \eqref{eq:score} and \eqref{eq:finite-score-sample}\;
 }
 Form $\widehat g_k$ and update $K_{k+1}$ using \eqref{eq:batch-update}\;
}
\KwOut{Linear feedback controller $\mu_{K_T}(x)=K_Tx$.}
\end{algorithm2e}
We next establish finite-time convergence and sample-complexity guarantees for the proposed first-order model-free PG algorithm for updating LQR gain $K$. 
\color{black}

\section{Convergence, Optimality, and Sample Complexity}\label{sec:convergence}
We now establish finite-time performance guarantees for Algorithm~\ref{alg:pg}. We first relate the optimality gap to the policy gradient through weak gradient dominance, then combine it with local smoothness and gradient-estimator conditions to derive convergence and sample-complexity bounds.

\subsection{Optimality}\label{sec:landscape}
Let $K\in\mathcal S$ be a stabilizing gain and recall that
$F_K=A+BK$ is the state-transition matrix under the linear feedback
law $u_j=Kx_j$. If the initial state has second moment
$M=\E[x_0x_0^\top]\succeq0$, then $x_j=F_K^jx_0$ and
$\E[x_jx_j^\top]=F_K^jM(F_K^j)^\top$. We now define the following series
\[
 S_K(M):=\sum_{j=0}^{\infty}F_K^jM(F_K^j)^\top
\]
for the sum of these state second moments. Note that this series converges because
$F_K$ is Schur stable.
%
%
Furthermore, we define the following
\begin{equation}\label{eq:gd-constant}
 C:=\frac{\|S_{K^*}(\Sigma_0)\|}
 {4\lambda_{\min}(R)\lambda_{\min}(W)^2},
 \quad
 \kappa:=\|\Sigma_0^{-1/2}B\Omega B^\top\Sigma_0^{-1/2}\|,
 \quad
 \beta:=s_\sigma\kappa C.
\end{equation}
%
%
%
Here $\kappa$ is the smallest scalar such that $B\Omega B^\top\preceq\kappa\Sigma_0$. Since $K^*$ minimizes the
classical LQR objective from every initial state $x$, we have
$x^\top P_{K^*}x\le x^\top P_Kx$ for all $x$ and $K\in\mathcal S$.
Thus $P_K-P_{K^*}\succeq0$, and \eqref{eq:cost} gives
\[
 \Delta(K)-\Delta(K^*)
 =s_\sigma\tr\!\big((P_K-P_{K^*})B\Omega B^\top\big)
 \le s_\sigma\kappa\big(J_\mu(K)-J_\mu^*\big).
\]
Hence the exploration-cost difference is at most $\beta$ on the
sublevel set $\mathcal L:=\{K\in\mathcal S:J_\mu(K)-J_\mu^*\le C\}$.
%
We now present the following global optimality result for LQR objective under exploration.
\begin{theorem}[Weak Gradient Dominance]\label{thm:gd}
Suppose Assumption \ref{as:plant} holds. Then, every $K\in\mathcal L$ satisfies
\begin{equation}\label{eq:wgd-squared}
 J_\pi(K)-J_\pi^*
 \le C\|\nabla J_\pi(K)\|_F^2+\beta.
\end{equation}
\end{theorem}
%
This result bounds the optimality gap $J_\pi(K)-J_\pi^*$ by the squared policy-gradient norm and the exploration-dependent residual $\beta$. We next analyze finite-time performance of Algorithm~\ref{alg:pg}.
%
%
\subsection{Finite-Time Convergence and Sample Complexity}
Let $\mathcal F_k$ denote the history before iteration $k$ of Algorithm~\ref{alg:pg}, and write $\E_k[\cdot]:=\E[\cdot\mid\mathcal F_k]$. We impose the following conditions on the estimates used in its gain update.

\begin{assumption}[Gradient Estimator]\label{as:gradient-estimator}
There exists $V>0$ such that, at every iteration $k$ with $K_k\in\mathcal S$, the estimate $\widehat g_k$ satisfies $\|\widehat g_k\|_F\le1$ almost surely and
\begin{equation}\label{eq:estimator-assumption}
 \E_k[\widehat g_k]=\nabla J_\pi(K_k),
 \qquad
 \E_k\!\left[\|\widehat g_k-\nabla J_\pi(K_k)\|_F^2\right]
 \le \frac{V}{N}.
\end{equation}
\end{assumption}
We can construct an unbiased infinite-horizon policy gradient estimator with finite-horizon trajectory samples in Algorithm~\ref{alg:pg} following \citep{rhee2015unbiased}. However, such a construction is beyond the scope of this work where we instead focus on a unified first-order model-free PG algorithm for learning the optimal LQR gain. Furthermore, the unit-norm gradient bound is imposed for technical convenience and requires normalization only when its norm is greater than 1. With these conditions in place, we now obtain the following linear-norm bound for the convergence analysis.
\begin{corollary}\label{cor:wgd}
Suppose Assumptions~\ref{as:plant} and~\ref{as:gradient-estimator} hold. Then, every stabilizing iterate $K_k$ of Algorithm~\ref{alg:pg} belongs to $\mathcal L$ and satisfies
\begin{equation}\label{eq:wgd}
 J_\pi(K_k)-J_\pi^*
 \le C\|\nabla J_\pi(K_k)\|_F+\beta.
\end{equation}
\end{corollary}
To analyze descent along the gain updates, we choose uniform local smoothness constants around $\mathcal L$. Since $\mathcal L$ is a compact subset of $\mathcal S$, its closed Frobenius-norm $r$-neighborhood $\mathcal N_r$ lies in $\mathcal S$ for sufficiently small $r>0$. Twice continuous differentiability of $J_\pi$ then ensures a finite bound $L>0$ on its Hessian operator norm over $\mathcal N_r$, with respect to the Frobenius norm.
\begin{theorem}[Finite-Time Convergence]\label{thm:convergence}
Suppose Assumptions~\ref{as:plant} and~\ref{as:gradient-estimator} hold. Run Algorithm~\ref{alg:pg} from $K_0\in\mathcal S$, and let
$l_0:=(J_\pi(K_0)-J_\pi^*-\beta)_+$ and
$R_{\zeta,N}:=C\sqrt{L\zeta V/N}$.
For a step size satisfying
\begin{equation}\label{eq:step-size}
 0<\zeta\le\min\{r,1/L\},
 \qquad
 \zeta\max\{l_0,R_{\zeta,N}\}\le C^2,
\end{equation}
all gains $K_0,\ldots,K_T$ are stabilizing, and
\begin{equation}\label{eq:rate}
 \E[J_\pi(K_T)]-J_\mu^*
 \le \Delta(K^*)+\beta
 +l_0\left(1-\frac{\zeta R_{\zeta,N}}{2C^2}\right)^T
 +R_{\zeta,N}.
\end{equation}
\end{theorem}
Here $\Delta(K^*)$ is the difference between the optimal objective values, that is $J_{\pi}^*-J_{\mu}^*$, and $\beta$ is the exploration residual in the gradient-dominance result. The remaining terms depend on the update count, step size, and sampling variance in Algorithm~\ref{alg:pg}. We next specify the step size and update count in Algorithm~\ref{alg:pg} to obtain the corresponding sample-complexity bound.

\begin{corollary}[Finite-Time Sample Complexity]\label{cor:samples}
Under the assumptions of Theorem~\ref{thm:convergence}, fix $\epsilon>0$ and run Algorithm~\ref{alg:pg} with step size
$0<\zeta\le\min\{r,L^{-1},C^2/\max\{l_0,\epsilon/2\},N\epsilon^2/(4C^2LV)\}$
and update count
\[
 T\ge\max\!\left\{1,\left\lceil\frac{4C^2}{\zeta\epsilon}
       \log\!\left(\max\{1,2l_0/\epsilon\}\right)
       \right\rceil\right\}.
\]
Then,
\begin{equation}\label{eq:epsilon-bound}
 \E[J_\pi(K_T)]-J_\mu^*
 \le\Delta(K^*)+\beta+\epsilon.
\end{equation}
For fixed $N$ and problem constants, taking the largest stated step size and the smallest stated update count requires
$NT=O(\epsilon^{-3}\log(1/\epsilon))$ gradient-estimator samples as $\epsilon\downarrow0$.
\end{corollary}
%
%
Since $J_\pi^*-J_\mu^*=\Delta(K^*)$, the corollary gives
$\E[J_\pi(K_T)]-J_\pi^*\le\beta+\epsilon$. By \eqref{eq:deploy},
the returned linear feedback controller therefore satisfies
$\E[J_\mu(K_T)]-J_\mu^*\le\beta+\epsilon$.

\section{Conclusion}
We developed a first-order model-free PG framework for undiscounted infinite-horizon LQR using a stochastic wrapper policy class centered at linear feedback controllers. 
We characterized the cost of decaying exploration and showed that minimizing the resulting objective recovers the unique optimal classical LQR gain. The wrapper policy score yields gain updates from sampled states, control inputs, and stage costs, without system identification or value-function approximation.
We further established weak gradient dominance and, under the stated estimator conditions, finite-time convergence and sample-complexity guarantees that explicitly account for exploration. Future work includes explicit unbiased gradient estimator design and faster convergence guarantees.

\appendix
\section{Proofs}\label{app:proofs}\label{app:scores}\label{app:convergence}
\begin{enumerate}[label=(\roman*),ref=\roman*,leftmargin=0pt,itemindent=2em,
 labelwidth=1.5em,labelsep=0.5em,itemsep=7pt,topsep=0pt]
\item \textbf{Proof of Theorem~\ref{thm:cost}.}
Fix $K\in\mathcal S$ and let $V_K(x)=x^\top P_Kx$. Under the wrapper
policy, $x_{t+1}=F_Kx_t+\sigma_tBz_t$. Since $z_t$ is centered and
independent of $x_t$, the cross terms in $c_t$ and $V_K(x_{t+1})$
vanish in conditional expectation. The Lyapunov equation
\eqref{eq:lyap} then gives
\begin{equation}
 \E[c_t+V_K(x_{t+1})-V_K(x_t)\mid x_t]
 =\sigma_t^2\tr\!\big((R+B^\top P_KB)\Omega\big).
\end{equation}
Taking expectations and summing from $t=0$ to $T-1$, we obtain
\begin{equation}
 \E\!\left[\sum_{t=0}^{T-1}c_t\right]+\E[V_K(x_T)]
 =J_\mu(K)+\tr\!\big((R+B^\top P_KB)\Omega\big)
   \sum_{t=0}^{T-1}\sigma_t^2,
\end{equation}
where $\E[V_K(x_0)]=J_\mu(K)$. Since $V_K\ge0$ and
$\sum_{t\ge0}\sigma_t^2=s_\sigma<\infty$, the expected accumulated
cost is bounded uniformly in $T$. Thus $\E[c_T]\to0$, and
$c_T\ge\lambda_{\min}(Q)\|x_T\|^2$ implies $\E[V_K(x_T)]\to0$.
Letting $T\to\infty$ and applying monotone convergence gives
\eqref{eq:cost}. Substituting $J_\mu(K)=\tr(P_K\Sigma_0)$ into
\eqref{eq:cost} and using cyclicity of the trace gives
\eqref{eq:weighted-cost}.
The Lyapunov operator $X\mapsto X-F_K^\top XF_K$ is invertible on
$\mathcal S$ by the Lyapunov series. The operator and $Q+K^\top RK$
are polynomial in $K$, so smoothness of inversion gives $P_K\in C^2$
on $\mathcal S$. Equation~\eqref{eq:weighted-cost} gives $J_\pi\in C^2$.
\hfill$\square$
\item \textbf{Proof of Theorem~\ref{thm:optimality}.}
Let $P_*:=P_{K^*}$ and $H_*:=R+B^\top P_*B\succ0$.
Completing the square in the classical Riccati equation
\citep{anderson2007} gives
$Q+K^\top RK+F_K^\top P_*F_K-P_*=D_K$, where
$D_K:=(K-K^*)^\top H_*(K-K^*)\succeq0$.
By \eqref{eq:lyap}, $P_K-P_*=D_K+F_K^\top(P_K-P_*)F_K$.

The Lyapunov series then gives
$P_K-P_*=\sum_{j\ge0}(F_K^j)^\top D_KF_K^j\succeq D_K$.
Using \eqref{eq:weighted-cost}, we obtain
\begin{equation}
 J_\pi(K)-J_\pi(K^*)
 =\tr\!\big((P_K-P_*)W\big)
 \ge\|H_*^{1/2}(K-K^*)W^{1/2}\|_F^2.
\end{equation}
Furthermore, since $H_*\succ0$ and $W\succ0$, the lower bound is positive whenever
$K\ne K^*$. Thus $K^*$ is the unique minimizer of $J_\pi$ over
$\mathcal S$. Further, evaluating \eqref{eq:cost} at $K^*$ gives
$J_\pi^*=J_\mu^*+\Delta(K^*)$. Hence, for every $K\in\mathcal S$,
\begin{equation}\label{app:eq:gap-decomposition}
 J_\pi(K)-J_\pi^*
 =J_\mu(K)-J_\mu^*+\Delta(K)-\Delta(K^*).
\end{equation}
Finally, since $P_K-P_*\succeq0$ and $B\Omega B^\top\succeq0$, the exploration
cost difference
$\Delta(K)-\Delta(K^*)=s_\sigma\tr\!\big((P_K-P_*)B\Omega B^\top\big)$
is nonnegative. This proves \eqref{eq:deploy}.
\hfill$\square$
\item \textbf{Proof of Theorem~\ref{thm:gd}.}
Let $E_K:=RK+B^\top P_KF_K$ and $\Sigma_*:=S_{K^*}(\Sigma_0)$.
By \eqref{eq:weighted-cost}, $J_\pi(K)$ is the classical LQR objective
with initial second moment $W$, up to a constant independent of $K$.
The LQR gradient formula \citep[Lemma~1]{fazel2018}, written for the
feedback convention $u=Kx$, gives $\nabla J_\pi(K)=2E_KS_K(W)$.
Since $S_K(W)\succeq W\succ0$, it follows that
$\|E_K\|_F\le\|\nabla J_\pi(K)\|_F/(2\lambda_{\min}(W))$.
Then, applying the upper bound in \citet[Supplement, Lemma~11]{fazel2018}
with initial second moment $\Sigma_0$ gives
\begin{equation}\label{eq:det-gap-gradient}
 J_\mu(K)-J_\mu^*
 \le \frac{\|\Sigma_*\|}{\lambda_{\min}(R)}\|E_K\|_F^2
 \le C\|\nabla J_\pi(K)\|_F^2,\qquad K\in\mathcal S.
\end{equation}
Finally, for $K\in\mathcal L$, the bound $\Delta(K)-\Delta(K^*)\le\beta$ and \eqref{eq:det-gap-gradient} give the desired
result using \eqref{app:eq:gap-decomposition}.
\hfill$\square$

\item \textbf{Proof of Corollary~\ref{cor:wgd}.}
At a stabilizing iterate $K_k$, Assumption~\ref{as:gradient-estimator}
and conditional Jensen's inequality give
\begin{equation}
 \|\nabla J_\pi(K_k)\|_F
 =\|\E_k[\widehat g_k]\|_F
 \le\E_k[\|\widehat g_k\|_F]\le1.
\end{equation}
Thus \eqref{eq:det-gap-gradient} gives $J_\mu(K_k)-J_\mu^*\le C$,
so $K_k\in\mathcal L$. Finally, applying Theorem~\ref{thm:gd}, we get
\eqref{eq:wgd}.
\hfill$\square$
\item \textbf{Proof of Theorem~\ref{thm:convergence}.}
Let $q:=\lambda_{\min}(Q)$ and $w:=\lambda_{\min}(\Sigma_0)$.
For $K\in\mathcal S$, the initial input cost and the accumulated
state cost give, respectively,
\begin{equation}
\begin{aligned}
 J_\mu(K)&\ge w\lambda_{\min}(R)\|K\|_F^2,\\
 J_\mu(K)&\ge qw\sum_{j\ge0}\|F_K^j\|_F^2
 \ge \frac{qw}{1-\rho(F_K)^2}.
\end{aligned}
\end{equation}
%
On $\mathcal L$, $J_\mu(K)\le J_\mu^*+C$, so the gains are bounded
and $\rho(F_K)$ is uniformly below one. Continuity of $\rho$ and $J_\mu$
therefore gives compactness of $\mathcal L$ in $\mathcal S$. Its closed
$r$-neighborhood $\mathcal N_r$ is compact and lies in $\mathcal S$
for small enough $r>0$. Theorem~\ref{thm:cost} gives a uniform Hessian
bound $L$ on $\mathcal N_r$.

If $K_k\in\mathcal S$, Corollary~\ref{cor:wgd} puts $K_k$ in $\mathcal L$,
and $\zeta\|\widehat g_k\|_F\le r$ almost surely keeps the update segment
in $\mathcal N_r\subset\mathcal S$. Induction from $K_0\in\mathcal S$
gives stability. Continuity of $J_\pi$ on $\mathcal N_r$ gives integrability.

Write $g_k:=\nabla J_\pi(K_k)$. Applying Taylor's inequality along the
update segment, taking conditional expectations, and using
$\E_k[\widehat g_k]=g_k$,
$\E_k[\|\widehat g_k\|_F^2]\le\|g_k\|_F^2+V/N$, and $L\zeta\le1$ gives
\begin{equation}
 \E_k[J_\pi(K_{k+1})]
 \le J_\pi(K_k)-\frac{\zeta}{2}\|g_k\|_F^2+\frac{L\zeta^2V}{2N}.
\end{equation}
Let $d_k:=\E[J_\pi(K_k)]-J_\pi^*-\beta$,
$a:=\zeta/(2C^2)$, and $b:=L\zeta^2V/(2N)$.
Corollary~\ref{cor:wgd} and Jensen's inequality for
$s\mapsto(s_+)^2$ give $(d_k)_+^2\le C^2\E[\|g_k\|_F^2]$. Hence
\begin{equation}\label{app:eq:scalar-recursion}
 d_{k+1}\le d_k-a(d_k)_+^2+b,\qquad
 (d_0)_+=l_0,\qquad \sqrt{b/a}=R_{\zeta,N}.
\end{equation}
To bound this recursion, we adapt the scalar comparison argument of \citet[Appendix~F]{montenegro2024}.
For $D\ge R_{\zeta,N}$ with $2a\max\{l_0,D\}\le1$, set
$u_0:=\max\{l_0,D\}$ and $u_{k+1}:=\Phi_D(u_k)$, where
$\Phi_D(u):=u-a(u_+)^2+aD^2$. The condition $2au_0\le1$ makes
$\Phi_D$ nondecreasing on $(-\infty,u_0]$. Since $\Phi_D(D)=D$ and
$\Phi_D(u)\le u$ for $u\ge D$, the sequence stays in $[D,u_0]$.
Also, $d_0\le u_0$ and $b\le aD^2$, so
\eqref{app:eq:scalar-recursion} gives $d_k\le u_k$ by induction.
Since $u_k\ge D$, we have
$u_{k+1}-D=(u_k-D)[1-a(u_k+D)]\le(1-aD)(u_k-D)$.
Iterating and using $u_0-D\le l_0$ yields
\begin{equation}\label{app:eq:comparison-bound}
 d_T\le D+l_0(1-aD)^T.
\end{equation}
By \eqref{eq:step-size}, we may take $D=R_{\zeta,N}$.
Adding $\beta+J_\pi^*-J_\mu^*=\beta+\Delta(K^*)$ gives \eqref{eq:rate}.
\hfill$\square$

\item \textbf{Proof of Corollary~\ref{cor:samples}.}
The prescribed step size gives $R_{\zeta,N}\le\epsilon/2$ and
$2a\max\{l_0,\epsilon/2\}\le1$. Taking $D=\epsilon/2$ in
\eqref{app:eq:comparison-bound} and using $1-x\le e^{-x}$ yields
\begin{equation}
 d_T\le\frac{\epsilon}{2}
 +l_0\exp\!\left(-\frac{\zeta\epsilon T}{4C^2}\right).
\end{equation}
The second term is at most $\epsilon/2$ directly if
$l_0\le\epsilon/2$, and by the prescribed lower bound on $T$ otherwise.
Thus $d_T\le\epsilon$, and adding $\beta+\Delta(K^*)$ gives
\eqref{eq:epsilon-bound}.

For fixed $N$ and problem constants, the largest permitted step size
is $\Theta(\epsilon^2)$ as $\epsilon\downarrow0$. The smallest
prescribed update count is then $O(\epsilon^{-3}\log(1/\epsilon))$
if $l_0>0$, and $T=1$ if $l_0=0$. Each update uses $N$ gradient
samples, giving the stated bound on $NT$.
\hfill$\square$
\end{enumerate}

\clearpage
\bibliography{references}
\end{document}


\maketitle
We provide a detailed version of the proofs of the main-paper appendix.
\section*{Supplementary Proofs}\label{supp:proofs}
Consider the stabilizable system $x_{t+1}=Ax_t+Bu_t$ with stage cost
$c_t=x_t^\top Qx_t+u_t^\top Ru_t$, where $Q,R\succ0$. The initial
state has finite second moment $\Sigma_0=\E[x_0x_0^\top]\succ0$.
For the linear feedback law $\mu_K(x)=Kx$, let $F_K=A+BK$ and
$Q_K=Q+K^\top RK$. The set of stabilizing gains is
$\mathcal S=\{K:\rho(F_K)<1\}$. The Gaussian wrapper $\pi_K$ samples inputs as
\begin{equation}\label{supp:eq:wrapper}
 u_t=Kx_t+\sigma_t z_t,\qquad \sigma_t=\frac{\sigma}{t+1},\qquad
 z_t\sim\mathcal N(0,\Omega).
\end{equation}
The $z_t$ are independent across time and independent of $x_0$.
The parameters $\sigma>0$ and $\Omega\succ0$ are fixed independently
of $K$. Write $M_B:=B\Omega B^\top$ and $s_\sigma:=\sum_{t\ge0}\sigma_t^2=\pi^2\sigma^2/6$.
The objectives $J_\mu(K)$ and $J_\pi(K)$ are the expected sums of
$c_t$ under $\mu_K$ and $\pi_K$, respectively.

\begin{enumerate}[label=(\roman*),ref=\roman*,leftmargin=0pt,itemindent=2em,
 labelwidth=1.5em,labelsep=0.5em,itemsep=7pt,topsep=0pt]

\item \textbf{Classical policy evaluation and the stabilizing domain.}
Under $\mu_K$, the state is $x_t=F_K^t x_0$. For Schur $F_K$,
geometric decay of its powers gives the convergent Lyapunov series
\begin{equation}\label{supp:eq:P-series}
 P_K=\sum_{j=0}^{\infty}(F_K^j)^\top Q_K F_K^j,\qquad
 J_\mu(K)=\tr(P_K\Sigma_0).
\end{equation}
Shifting the series gives the policy-evaluation equation
\begin{equation}\label{supp:eq:lyap}
 P_K=Q_K+F_K^\top P_K F_K .
\end{equation}
If two matrices solve this equation, their difference $X$ satisfies
$X=(F_K^j)^\top X F_K^j$ for every $j$. The right-hand side tends to
zero, proving uniqueness. Set $q:=\lambda_{\min}(Q)$ and
$w:=\lambda_{\min}(\Sigma_0)$. The accumulated state cost gives
\begin{equation}\label{supp:eq:stability-lower}
 J_\mu(K)\ge qw\sum_{j\ge0}\|F_K^j\|_F^2
 \ge qw\sum_{j\ge0}\rho(F_K)^{2j}.
\end{equation}
If $\rho(F_K)\ge1$, the right-hand side diverges. Thus $J_\mu(K)$
is finite exactly on $\mathcal S$. Under stabilizability and
$Q,R\succ0$, classical LQR theory gives the stabilizing Riccati
solution and the optimal gain $K^*$ \citep{anderson2007}.
Write $J_\mu^*:=J_\mu(K^*)$ and $J_\pi^*:=\inf_{K\in\mathcal S}J_\pi(K)$.
\hfill$\square$

\item \textbf{Proof of \MainResult{thm:cost}{Theorem}{the cost of exploration theorem}.}
Fix $K\in\mathcal S$ and let $V_K(x)=x^\top P_Kx$. Under the wrapper
policy, $x_{t+1}=F_Kx_t+\sigma_tBz_t$. Since $z_t$ is centered and
independent of $x_t$, the cross terms in $c_t$ and $V_K(x_{t+1})$
vanish in conditional expectation. The Lyapunov equation
\eqref{supp:eq:lyap} then gives
\begin{equation}
 \E[c_t+V_K(x_{t+1})-V_K(x_t)\mid x_t]
 =\sigma_t^2\tr\!\big((R+B^\top P_KB)\Omega\big).
\end{equation}
Taking expectations and summing from $t=0$ to $T-1$, we obtain
\begin{equation}
 \E\!\left[\sum_{t=0}^{T-1}c_t\right]+\E[V_K(x_T)]
 =J_\mu(K)+\tr\!\big((R+B^\top P_KB)\Omega\big)
   \sum_{t=0}^{T-1}\sigma_t^2,
\end{equation}
Here $\E[V_K(x_0)]=J_\mu(K)$. Because $V_K\ge0$ and
$s_\sigma<\infty$, the sum of the nonnegative expected stage costs is
finite. In particular, $q\E[\|x_T\|^2]\le\E[c_T]\to0$, so
$\E[V_K(x_T)]\to0$.
Letting $T\to\infty$ and applying monotone convergence gives
\begin{equation}\label{supp:eq:cost-decomposition}
 J_\pi(K)=J_\mu(K)+\Delta(K),\qquad
 \Delta(K):=s_\sigma\tr\!\big((R+B^\top P_KB)\Omega\big).
\end{equation}
Using $J_\mu(K)=\tr(P_K\Sigma_0)$ and cyclicity of the trace gives
\begin{equation}\label{supp:eq:weighted-cost}
 J_\pi(K)=\tr(P_KW)+s_\sigma\tr(R\Omega),
 \qquad W:=\Sigma_0+s_\sigma M_B\succ0.
\end{equation}

For $K\in\mathcal S$, the Lyapunov operator
$X\mapsto X-F_K^\top XF_K$ is invertible, with inverse given by the
Lyapunov series. Both the operator and $Q+K^\top RK$ depend
polynomially on $K$. Smoothness of matrix inversion therefore implies
that $P_K$ is twice continuously differentiable on $\mathcal S$,
and \eqref{supp:eq:weighted-cost} gives the same regularity for $J_\pi$.
\hfill$\square$

\item \textbf{Proof of \MainResult{thm:optimality}{Theorem}{the optimal gain theorem}.}
Let $P_*:=P_{K^*}$ and $H_*:=R+B^\top P_*B\succ0$.
Completing the square in the classical Riccati equation
\citep{anderson2007} gives
$Q+K^\top RK+F_K^\top P_*F_K-P_*=D_K$, where
$D_K:=(K-K^*)^\top H_*(K-K^*)\succeq0$.
Substituting this identity into \eqref{supp:eq:lyap} yields
\begin{equation}
 P_K-P_*=D_K+F_K^\top(P_K-P_*)F_K.
\end{equation}
For $K\in\mathcal S$, the Lyapunov series gives
$P_K-P_*=\sum_{j\ge0}(F_K^j)^\top D_KF_K^j\succeq D_K$.
Using \eqref{supp:eq:weighted-cost}, we obtain
\begin{equation}
 J_\pi(K)-J_\pi(K^*)
 =\tr\!\big((P_K-P_*)W\big)
 \ge\|H_*^{1/2}(K-K^*)W^{1/2}\|_F^2.
\end{equation}
Since $H_*\succ0$ and $W\succ0$, the lower bound is positive whenever
$K\ne K^*$. Thus $K^*$ is the unique minimizer of $J_\pi$ over
$\mathcal S$. Evaluating \eqref{supp:eq:cost-decomposition} at $K^*$ gives
$J_\pi^*=J_\mu^*+\Delta(K^*)$. Hence, for every $K\in\mathcal S$,
\begin{equation}\label{supp:eq:gap-decomposition}
 J_\pi(K)-J_\pi^*
 =J_\mu(K)-J_\mu^*+\Delta(K)-\Delta(K^*).
\end{equation}
Since $P_K-P_*\succeq0$ and $B\Omega B^\top\succeq0$, the exploration
cost difference
$\Delta(K)-\Delta(K^*)=s_\sigma\tr\!\big((P_K-P_*)B\Omega B^\top\big)$
is nonnegative. Hence
$J_\mu(K)-J_\mu^*\le J_\pi(K)-J_\pi^*$.
\hfill$\square$

\item \textbf{Proof of \MainResult{thm:gd}{Theorem}{the weak gradient dominance theorem}.}
For $M\succeq0$, let $S_K(M):=\sum_{j\ge0}F_K^jM(F_K^j)^\top$ be the
sum of state second moments under $u_t=Kx_t$ with initial second moment
$M$. Let $E_K:=RK+B^\top P_KF_K$ and $\Sigma_*:=S_{K^*}(\Sigma_0)$.
Define
\begin{equation}
\begin{aligned}
 C&:=\frac{\|\Sigma_*\|}{4\lambda_{\min}(R)\lambda_{\min}(W)^2},\\
 \kappa&:=\|\Sigma_0^{-1/2}M_B\Sigma_0^{-1/2}\|,
 \qquad \beta:=s_\sigma\kappa C .
\end{aligned}
\end{equation}
Let $\mathcal L:=\{K\in\mathcal S:J_\mu(K)-J_\mu^*\le C\}$.
By \eqref{supp:eq:weighted-cost}, $J_\pi(K)$ is the classical LQR objective
with initial second moment $W$, up to a constant independent of $K$.
The LQR gradient formula \citep[Lemma~1]{fazel2018}, written for the
feedback convention $u=Kx$, gives $\nabla J_\pi(K)=2E_KS_K(W)$.
Since $S_K(W)\succeq W\succ0$, it follows that
$\|E_K\|_F\le\|\nabla J_\pi(K)\|_F/(2\lambda_{\min}(W))$.
Applying the upper bound in \citet[Supplement, Lemma~11]{fazel2018}
with initial second moment $\Sigma_0$ gives
\begin{equation}\label{supp:eq:det-gap-gradient}
 J_\mu(K)-J_\mu^*
 \le \frac{\|\Sigma_*\|}{\lambda_{\min}(R)}\|E_K\|_F^2
 \le C\|\nabla J_\pi(K)\|_F^2,\qquad K\in\mathcal S.
\end{equation}

By definition of $\kappa$, $B\Omega B^\top\preceq\kappa\Sigma_0$.
Taking the trace against $P_K-P_*\succeq0$ gives, for $K\in\mathcal L$,
\begin{equation}
 \Delta(K)-\Delta(K^*)
 \le s_\sigma\kappa\big(J_\mu(K)-J_\mu^*\big)
 \le\beta.
\end{equation}
Combining these bounds in \eqref{supp:eq:gap-decomposition} gives
$J_\pi(K)-J_\pi^*\le C\|\nabla J_\pi(K)\|_F^2+\beta$ for $K\in\mathcal L$.
\hfill$\square$

\item \textbf{Proof of \MainResult{cor:wgd}{Corollary}{the corollary on stabilizing iterates}.}
Write $\E_k$ for expectation conditional on the history before
iteration $k$. At each stabilizing iterate, the gradient-estimator
assumption in the main paper requires
\begin{equation}\label{supp:eq:estimator-conditions}
\begin{gathered}
 \E_k[\widehat g_k]=\nabla J_\pi(K_k),\qquad
 \|\widehat g_k\|_F\le1\quad\text{almost surely},\\
 \E_k\!\left[\|\widehat g_k-\nabla J_\pi(K_k)\|_F^2\right]\le V/N .
\end{gathered}
\end{equation}
Conditional Jensen's inequality gives
\begin{equation}
 \|\nabla J_\pi(K_k)\|_F
 =\|\E_k[\widehat g_k]\|_F
 \le\E_k[\|\widehat g_k\|_F]\le1.
\end{equation}
Thus \eqref{supp:eq:det-gap-gradient} gives $J_\mu(K_k)-J_\mu^*\le C$,
so $K_k\in\mathcal L$. Since $\|\nabla J_\pi(K_k)\|_F\le1$,
the preceding bound gives
$J_\pi(K_k)-J_\pi^*\le C\|\nabla J_\pi(K_k)\|_F+\beta$.
\hfill$\square$

\item \textbf{Proof of \MainResult{thm:convergence}{Theorem}{the finite-time convergence theorem}.}
For $K\in\mathcal S$, the initial input cost and
\eqref{supp:eq:stability-lower} give, respectively,
\begin{equation}
\begin{aligned}
 J_\mu(K)&\ge w\lambda_{\min}(R)\|K\|_F^2,\\
 J_\mu(K)&\ge qw\sum_{j\ge0}\|F_K^j\|_F^2
 \ge \frac{qw}{1-\rho(F_K)^2}.
\end{aligned}
\end{equation}
On $\mathcal L$, $J_\mu(K)\le J_\mu^*+C$, so the gains are bounded
and $\rho(F_K)$ is uniformly below one. Continuity of $\rho$ puts
every limit of gains in $\mathcal L$ in $\mathcal S$, where continuity
of $J_\mu$ preserves the cost bound. Thus $\mathcal L$ is compact.
For sufficiently small $r>0$, its closed Frobenius-norm
$r$-neighborhood $\mathcal N_r$ is compact and contained in
$\mathcal S$. The continuous Hessian of $J_\pi$ has a uniform bound
$L>0$ there.

The update is $K_{k+1}=K_k-\zeta\widehat g_k$. Set
$l_0:=(J_\pi(K_0)-J_\pi^*-\beta)_+$ and
$R_{\zeta,N}:=C\sqrt{L\zeta V/N}$. The step size satisfies
\begin{equation}\label{supp:eq:step-size}
 0<\zeta\le\min\{r,1/L\},\qquad
 \zeta\max\{l_0,R_{\zeta,N}\}\le C^2.
\end{equation}
If $K_k\in\mathcal S$, the preceding corollary gives $K_k\in\mathcal L$.
By \eqref{supp:eq:estimator-conditions} and \eqref{supp:eq:step-size},
$\zeta\|\widehat g_k\|_F\le r$ almost surely. The update segment
therefore lies in $\mathcal N_r$, so $K_{k+1}\in\mathcal S$. Induction
from $K_0\in\mathcal S$ proves stability of all iterates. The objective
values are integrable because $J_\pi$ is continuous on the compact
set $\mathcal N_r$.

Write $g_k:=\nabla J_\pi(K_k)$. Applying Taylor's inequality along the
update segment, taking conditional expectations, and using
$\E_k[\widehat g_k]=g_k$,
$\E_k[\|\widehat g_k\|_F^2]\le\|g_k\|_F^2+V/N$, and $L\zeta\le1$ gives
\begin{equation}
 \E_k[J_\pi(K_{k+1})]
 \le J_\pi(K_k)-\frac{\zeta}{2}\|g_k\|_F^2+\frac{L\zeta^2V}{2N}.
\end{equation}
Let $d_k:=\E[J_\pi(K_k)]-J_\pi^*-\beta$,
$a:=\zeta/(2C^2)$, and $b:=L\zeta^2V/(2N)$.
The preceding corollary and Jensen's inequality for
$s\mapsto(s_+)^2$ give $(d_k)_+^2\le C^2\E[\|g_k\|_F^2]$. Hence
\begin{equation}\label{supp:eq:scalar-recursion}
 d_{k+1}\le d_k-a(d_k)_+^2+b,\qquad
 (d_0)_+=l_0,\qquad \sqrt{b/a}=R_{\zeta,N}.
\end{equation}

To bound this recursion, we adapt the scalar comparison argument of
\citet[Appendix~F]{montenegro2024}. 

For $D\ge R_{\zeta,N}$ with
$2a\max\{l_0,D\}\le1$, let $u_0:=\max\{l_0,D\}$ and
$u_{k+1}:=\Phi_D(u_k)$, where $\Phi_D(u):=u-a(u_+)^2+aD^2$.
The map is nondecreasing on $(-\infty,u_0]$ because $2au_0\le1$.
Since $\Phi_D(D)=D$ and $\Phi_D(u)\le u$ for $u\ge D$, the iterates
$u_k$ stay in $[D,u_0]$. 

Also, $d_0\le u_0$ and $b\le aD^2$.
Monotonicity and \eqref{supp:eq:scalar-recursion} then give $d_k\le u_k$
by induction. On $[D,u_0]$,
\begin{equation*}
 u_{k+1}-D=(u_k-D)\big[1-a(u_k+D)\big]\le(1-aD)(u_k-D).
\end{equation*}
Iterating and using $u_0-D\le l_0$ yields
\begin{equation}\label{supp:eq:comparison-bound}
 d_T\le D+l_0(1-aD)^T.
\end{equation}
By \eqref{supp:eq:step-size}, we may take $D=R_{\zeta,N}$.
Adding $\beta+J_\pi^*-J_\mu^*=\beta+\Delta(K^*)$ yields
\begin{equation}
 \E[J_\pi(K_T)]-J_\mu^*
 \le\Delta(K^*)+\beta
 +l_0\left(1-\frac{\zeta R_{\zeta,N}}{2C^2}\right)^T+R_{\zeta,N}.
\end{equation}
\hfill$\square$

\item \textbf{Proof of \MainResult{cor:samples}{Corollary}{the finite-time sample complexity corollary}.}
For $\epsilon>0$, the corollary specifies the step size
$0<\zeta\le\min\{r,L^{-1},C^2/\max\{l_0,\epsilon/2\},N\epsilon^2/(4C^2LV)\}$.
These restrictions give $R_{\zeta,N}\le\epsilon/2$ and
$2a\max\{l_0,\epsilon/2\}\le1$. Taking $D=\epsilon/2$ in
\eqref{supp:eq:comparison-bound} and using $1-x\le e^{-x}$ yields
\begin{equation}
 d_T\le\frac{\epsilon}{2}
 +l_0\exp\!\left(-\frac{\zeta\epsilon T}{4C^2}\right).
\end{equation}
If $l_0\le\epsilon/2$, the second term is at most $\epsilon/2$.
If $l_0>\epsilon/2$, the iteration bound
$\zeta\epsilon T/(4C^2)\ge\log(2l_0/\epsilon)$ gives the same conclusion.
Thus $d_T\le\epsilon$, or equivalently
$\E[J_\pi(K_T)]-J_\pi^*\le\beta+\epsilon$. The identity
$J_\pi^*=J_\mu^*+\Delta(K^*)$ gives the stated bound, and
$J_\mu(K_T)-J_\mu^*\le J_\pi(K_T)-J_\pi^*$ gives
$\E[J_\mu(K_T)]-J_\mu^*\le\beta+\epsilon$.

For fixed $N$ and problem constants, the largest permitted step size
is $\Theta(\epsilon^2)$ as $\epsilon\downarrow0$. The smallest
prescribed update count is then $O(\epsilon^{-3}\log(1/\epsilon))$
if $l_0>0$, and $T=1$ if $l_0=0$. Each update uses $N$ gradient
samples, giving the stated bound on $NT$.
\hfill$\square$

\end{enumerate}

\bibliography{references}